\documentclass[12pt]{article}
\newcommand{\be}{\begin{equation}}
\newcommand{\ee}{\end{equation}}
\newcommand{\bqa}{\begin{eqnarray}}
\newcommand{\eqa}{\end{eqnarray}}

\begin{document}
\begin{center}
{\LARGE Interpretations for the $X(4160)$ observed in the double charm production at B factories }\\[0.8cm]
{\large Kuang-Ta Chao$^{(a,b)}$}\\[0.5cm]
{\footnotesize (a)~Department of Physics, Peking University,
Beijing 100871, China}\\
{\footnotesize (b)~Center for High Energy Physics, Peking
University, Beijing 100871, China}
\end{center}
\vspace{0.5cm}

\begin{abstract}
Belle Collaboration has recently observed a new state, the X(4160),
in the process of double charm production $e^+e^-\to J/\psi+X(4160)$
followed by $X(4160)\to D^*\bar{D^*}$. We discuss possible
interpretations for the X(4160) based on the NRQCD calculations and
the potential model estimates for the charmonium spectrum. We first
focus on the D-wave spin-singlet $2^{-+}$ charmonium $^1D_2(2D)$,
which is estimated to have a small production rate of about 5\% of
that for $e^+e^-\to J/\psi+\eta_c(1S)$, and therefore is
incompatible with the observed data for X(4160). We then discuss the
possibility that the X(4160) is the known $J^{PC}=1^{--}$ charmonium
state $\psi(4160)$, which can be produced via two photon
fragmentation, but the production rate is much smaller than observed
for $e^+e^-\to J/\psi+X(4160)$. In contrast to above two
possibilities, the $\eta_c(4S)$ assignment is a likely one, which is
supported by the observed relatively large production rate and
non-observation of $D\bar D$ decay of X(4160), but we have to
understand why $\eta_c(4S)$ has such a low mass, which deserves
further studies. The P-wave excited state $\chi_{c0}(3P)$ is also an
interesting candidate, if the observed broad peak around 3.8-3.9~GeV
in the recoil mass of $D\bar D$ against $J/\psi$ in
$e^++e^-\rightarrow J/\psi+D\bar D$ is due to the $\chi_{c0}(2P)$
state. Measurements of production angular distributions will be
helpful to distinguish between $\eta_c(4S)$ and $\chi_{c0}(3P)$
assignments.  Production mechanisms in nonrelativistic QCD are
emphasized.
\vspace{1cm}\\
\end{abstract}


\vspace{1cm}

Using a data sample of 693~fb$^{-1}$ collected around the
$\Upsilon(4S)$ with the Belle detector at the KEKB $e^+e^-$ storage
rings, very recently the Belle Collaboration has reported some new
results for double charmonium production in $e^+e^-$ annihilation at
$\sqrt{s}=10.6$~GeV\cite{belle0707}. In the measured processes
$e^+e^-\to J/\psi D^{(*)}\overline{D^{(*)}}$, a new resonance state,
called the X(4160), is observed with a significance of 5.1~$\sigma$
in $e^+e^-\to J/\psi X(4160)$ followed by $X(4160)\to D^*\bar{D^*}$.
As a hadronic resonance, the X(4160) has the following mass and
width\cite{belle0707} \bqa
  M=4156^{+25}_{-20}\pm {15}~MeV, ~~~~~~\Gamma=139^{+111}_{-61}\pm
  21~MeV,
\eqa and production cross section \be \sigma(e^+e^-\to J/\psi
X(4160))B_{D^*\bar{D^*}}=(24.7^{+12.8}_{-8.3}\pm 5.0)~fb, \ee which
is large and comparable to the observed cross sections for
$e^++e^-\rightarrow J/\psi+\eta_c(1S)$ and $e^++e^-\rightarrow
J/\psi+\chi_{c0}(1P)$. Although at present the data for the X(4160)
are still preliminary and more data are apparently needed to
identify the nature of this new state, it is worthwhile to discuss
its possible assignments, especially in view of the great potential
of finding new particles, e.g. the $\eta_c(2S)$ and the X(3940)(for
recent reviews on new hadrons with heavy quarks, see, e.g.
\cite{swanson06, zhu07}) in the $e^+e^-$ annihilation processes at B
factories. It is also interesting to study the production mechanisms
of those charmonium or charmonium-like states in the double
charmonium production processes in $e^+e^-$ annihilation, since the
theoretical understanding for the double charm production is very
intriguing but not totally conclusive.

In the following, we discuss some possible interpretations for the
X(4160), in connection with the double charmonium production
problem.

In general, in the process $e^+e^-\to J/\psi D^{*}\bar{D^{*}}$ the
$D^{*}\bar{D^{*}}$ system can have charge parity either C=+ (if
$e^+e^-$ annihilated into one photon) or C=- (if $e^+e^-$
annihilated into two photons). In the case of C=+, the X(4160) can
have $J^{PC}=0^{++}, 0^{-+}, 1^{-+}, 2^{-+}, 1^{++}, 2^{++}, ...$;
while in the case of C=-, it will have $J^{PC}=1^{--}, 2^{--},
1^{+-}, 2^{+-}, ...$ Because the two photon processes are relatively
suppressed by an additional electromagnetic coupling constant
$\alpha=1/137$, at the B factory energy $\sqrt{s}=10.6$~GeV the one
photon processes usually have larger rates, and should therefore be
considered firstly.

\vspace{0.5cm}

I. The assignment that the X(4160) is the D-wave spin-singlet
charmonium state $^1D_2(2D)$ is disfavored by the too small
theoretical rate of production in $e^+e^-$ annihilation.

This 2D state has quantum numbers $J^{PC}=2^{-+}$. There are some
arguments which could be in favor of this interpretation.

First, the observed X(4160) has the same mass as the $\psi(4160)$
(see the Particle Physics Booklet \cite{pdg}), which is known to be
the good candidate of the D-wave spin-triplet charmonium state
$^3D_1(2D)$ with $J^{PC}=1^{--}$,  and \bqa M=4153\pm 3~MeV,~~~~~
\Gamma=103\pm 8~MeV,~~~~ \Gamma_{ee}=0.83\pm 0.07~KeV. \eqa The
hyperfine splitting between the center of mass of $^3D_J(2D)$ states
and the $^1D_2(2D)$ is expected to be vanishing if the short-range
spin-dependent forces are due to one-gluon exchange, and the
fine-splittings between $^3D_J(2D)$ states should be a few tens MeV.
Therefore the observed mass of the X(4160), $M=4156^{+25}_{-20}\pm
{15}~MeV$, could be compatible with the D-wave spin-singlet
charmonium state $^1D_2(2D)$, and roughly speaking, is in agreement
with the potential model predictions (see, e.g.
\cite{Eichten:1979ms,Godfrey:1985xj,BGS0505002}). (Note, however,
that the S-D mixing effects should be considered due to a sizable
leptonic width observed for the $\psi(4160)$.)

Second, for the $J^{PC}=2^{-+}$ 2D charmonium state, the decay to
$D\bar D$ is forbidden, while decays to $D^*\bar{D^*}$ and
$D^*\bar{D}+c.c.$ are allowed (also including the $D_s^{*}\bar D_s$
mesons). Therefore, the decay $X(4160)\to D^*\bar{D^*}$ could be
substantial.
In ref.\cite{BGS0505002} the calculated decay rate of $2^{-+}(2D)\to
D^*\bar{D^*}$ is nearly equal to that of $2^{-+}(2D)\to
D^*\bar{D}+c.c.$. (However, the sensitivity to the model and
parameters need to be further investigated.)

However, the main problem for this assignment is that the D-wave
state is expected to have a much smaller production rate than the
S-wave state such as the $\eta_c$ in double charmonium production.
The double charmonium production in $e^+e^-$ annihilation at B
factories has been studied in the framework of nonrelativistic QCD
(NRQCD)\cite{bl,liu,liu04,h}, in which the charmonium states are
treated as nonrelativistic bound systems, and the production rates
can be factorized into the short distance part, which can be
calculated in perturbative QCD, and the long distance part, which
can be related to the wavefunction of the charmonium.
Experimentally, double charmonium production processes
$e^++e^-\rightarrow
J/\psi+\eta_c(1S)(\eta_c(2S),\chi_{c0}(1P),X(3940))$ have been
observed by Belle and BaBar \cite{Abe:2002rb,Pakhlov,BaBar:2005},
but the cross sections are larger than the leading order (LO) NRQCD
calculations by almost an order of magnitude\cite{bl,liu,h} (Note
that the  numerical results can be somewhat different when taking
different parameters e.g. in \cite{bl} and in \cite{liu} but the
physical conclusion is the same). The next to leading order (NLO)
QCD radiative corrections are found to be very significant to
increase the cross section of $e^++e^-\rightarrow
J/\psi+\eta_c(1S)$\cite{zhang}. Moreover, the relativistic
corrections further increase this cross section\cite{bodwin,he}. As
a result, the calculated cross section of $e^++e^-\rightarrow
J/\psi+\eta_c(1S)$ with both NLO radiative and relativistic
corrections in NRQCD may reach the lower bound of the experimental
values, and could resolve the problem.

Other approaches including the light-cone methods are also discussed
in the literature to resolve the discrepancy between experimental
data and theory for the double charmonium production\cite{ji}.

It is interesting to point out that although the LO results in NRQCD
for the double charmonium production cross sections in $e^++e^-$
annihilation are much smaller than data, the predicted relative
rates seem to be consistent with the measured values. For instance,
the predicted cross sections for $J/\psi+\eta_c(1S)$ and
$J/\psi+\chi_{c0}(1P)$ are comparable and much larger than that for
$J/\psi+\chi_{c1}(1P)$ and $J/\psi+\chi_{c2}(1P)$. These ratios of
LO cross sections are indeed compatible with data. In the following,
we will use the calculated LO result for $e^++e^-\rightarrow J/\psi+
^1D_2$ and compare it with that for $e^++e^-\rightarrow
J/\psi+\eta_c$, to make some predictions.

To the leading order in NRQCD with QED contribution included, the
cross section for ${e^++e^-\rightarrow \gamma^{*} \rightarrow
\psi(nS)+^1D_2(mD)}$ process can be expressed as\cite{liu04}
\begin{eqnarray}
\label{sssE} &&\sigma(e^+(p_1)+e^-(p_2)\rightarrow
\psi(p_3)+^1D_2(p_4))=\nonumber\\
&&\frac{5\alpha^2|R_{S}(0)|^2|R^{\prime\prime}_{D}(0)|^2\sqrt{s^2-2s(m_3^2+m_4^2)+(m_3^2-m_4^2)^2}
}{192 m_c^2 \pi s^2}\int^1_{-1}|\bar{M}|^2 dx,
\end{eqnarray}
and
\begin{eqnarray}
\label{j1d2E} {\mid \bar{M}_{J/\psi ^1D_2} \mid}^2
&=&\frac{4096\pi^2(s-16m_c^2)^3(32\alpha m_c^2+96\alpha_s m_c^2+3s
\alpha )^2(x^2+1)}{243m_c^6 s^6} ,
\end{eqnarray}
where $x=\cos \theta$, and $\theta$ is the angle between the beam
axis ($\overrightarrow{p_{1}}$) and the $J/\psi$ momentum ($
\overrightarrow{p_{3}}$).

As in\cite{liu,liu04}, we take following parameters:
$\sqrt{s}=10.6{\rm GeV},~~m_c=1.5{\rm GeV},~ m_3=m_4=2m_c$ (in the
nonrelativistic limit), $\alpha_s=0.26$, and the wave functions at
the origin are taken from a potential model calculation (see e.g.
the QCD (BT) model in Ref.\cite{wf}): $|R_{1S}(0)|^2 =0.810{\rm
GeV}^3$, and $|R_{1D}''(0)|^2=0.015 {\rm GeV^7}$,
$|R_{2D}''(0)|^2=0.024 {\rm GeV^7}$.

For the $J/\psi+ ^1D_2(1D)$ (see\cite{liu04}) and  $J/\psi+
^1D_2(2D)$ production, we then get the angular distributions
(differential cross sections) and cross sections, which are shown in
Table I,
where $\theta$ is the angle between the incident beam and the
$J/\psi$, and the numbers with (without) square brackets mean the
cross sections without QED (with QED) contributions. These cross
sections are much smaller than that predicted for the $J/\psi+
\eta_c(1S)$ production\cite{liu,liu04}, which is also listed in
Table I.
We see that the cross section for $J/\psi+ ^1D_2(2D)$ production is
predicted to be only about 5\% of that of $J/\psi+\eta_c$, in
contrast to the observed production cross section for the
$J/\psi+X(4160)$ shown in eq.(2), which is comparable to that of
$J/\psi+\eta_c$~\cite{Abe:2002rb,Pakhlov,BaBar:2005}.


In principle, we could also detect the $^1D_2(1D)$ charmonium, which
should lie around 3.8~GeV, in the $e^++e^-\rightarrow J/\psi+
^1D_2(1D)$ process. However, the main decay modes of $^1D_2(1D)$
should be decays to light hadrons via intermediate gluons, since the
$^1D_2(1D)$ is expected to lie below the $D^*\bar D$ threshold.
Without a dominant exclusive decay channel like $D^*\bar{D^*}$ or
$D^*\bar{D}+c.c.$, it will be even more difficult to detect this
charmonium state especially when the production cross section is
small.

To sum up, although the $^1D_2(2D)$ charmonium could be a possible
assignment for the X(4160), the predicted small production rate for
$e^++e^-\rightarrow J/\psi+ ^1D_2(2D)$ makes this assignment very
unlikely. Despite of the existing uncertainties in the theoretical
calculation (e.g., the chosen parameters, and high order
corrections), this conclusion should hold,  since the small number
of 5\% for the ratio of $J/\psi+ ^1D_2(2D)$ production cross section
to that of $J/\psi+\eta_c$  can not be enhanced to close to the
observed value (about 1) by changing the parameters or including the
NLO QCD corrections (Note that the NLO QCD correction to the
$J/\psi+\eta_c$ increases this production rate by a factor of about
2\cite{zhang}).

\vspace{0.5cm}

II. The possibility that the X(4160) is the known $J^{PC}=1^{--}$
charmonium state $\psi(4160)$ should be ruled out.

The $\psi(4160)$ is in the same mass region as the newly observed
X(4160), and their widths are also comparable. Moreover, the
$\psi(4160)$ can also decay to $D^*\bar{D^*}$. However, the process
$e^+e^-\to J/\psi+ \psi(4160)$ can only proceed through $e^+e^-$
annihilation into two photons due to the conserved charge parities.

In fact, the two-photon process was first studied for $e^+e^-\to
2\gamma^*\to J/\psi+ J/\psi$ in ref.\cite{cc}, and it was found that
the production rate is comparable to or even larger than that the
one-photon process $e^+e^-\to \gamma^*\to J/\psi+ \eta_c$ in the
leading order calculation\cite{cc}. Moreover, for the inclusive
double charm production process $e^+e^-\to J/\psi+ c\bar c$, the
two-photon process $e^+e^-\to 2\gamma^*\to J/\psi+ c\bar c$ will
prevail over the one-photon process $e^+e^-\to \gamma^*\to J/\psi+
c\bar c$ when $\sqrt{s}$ becomes larger than 20~GeV\cite{liuR}. This
is because, in these two-photon fragmentation processes the
virtualities of the photons are only about $4m_c^2$, which is much
smaller than the virtuality $s$ in the one-photon process.

However, because the $\psi(4160)$ is expected to be a D-wave
($^3D_1(2D)$) dominated charmonium state (with possibly some
$^3S_1(3S)$ admixture), its coupling to the photon is suppressed by
the factor $|sin\phi R_{3S}(0)-cos\phi
\frac{5}{2\sqrt{2}m_{c}^{2}}R_{2D}^{''}(0)|^{2}$, compared with
$|R_{1S}(0)|^2$ for the $J/\psi$. Here, we have assumed that the
$\psi(4160)$ is a mixture of the $^3D_1(2D)$ and $^3S_1(3S)$ states
with $\phi$ being the mixing angle:
\begin{eqnarray}
&&|\psi(4040)\rangle=|3{}^3{\rm S}_1\rangle \cos\phi+
 |2{}^3{\rm D}_1\rangle\sin\phi,\\
&&|\psi(4160)\rangle=-
 |3{}^3{\rm S}_1\rangle\sin\phi+ |2{}^3{\rm
D}_1\rangle\cos\phi.
\end{eqnarray}
The above expression is only a very rough approximation, since
admixtures with the charmed meson pairs due to coupled channel
effects and with other S-wave states are all ignored. With this
simple assumption we get leptonic decay widths for the $\psi(4040)$
and $\psi(4160)$:
\begin{eqnarray}
\label{mixing:ee1} \Gamma(\psi(4040)\rightarrow e^{+}e^{-}) =
4\alpha^{2}e_{c}^{2}\frac{|cos\phi R_{3S}(0)+sin\phi
\frac{5}{2\sqrt{2}m_{c}^{2}}R_{2D}^{''}(0)|^{2}}{(2m_c)^{2}}\,,\\
\label{mixing:ee2} \Gamma(\psi(4160)\rightarrow e^{+}e^{-}) =
4\alpha^{2}e_{c}^{2}\frac{|sin\phi R_{3S}(0)-cos\phi
\frac{5}{2\sqrt{2}m_{c}^{2}}R_{2D}^{''}(0)|^{2}}{(2m_c)^{2}}\,.
\end{eqnarray}
Using the experimental values $\Gamma_{ee}(\psi(4040))=0.86\pm
$0.07~KeV and $\Gamma_{ee}(\psi(4160))=0.83\pm $0.07~KeV, and $|
R_{3S}(0)|^2=0.455~GeV^3$, $|R_{2D}^{''}(0)|^{2}=0.024~GeV^7$, we
get the mixing angle from the ratio of these two leptonic widths:
\begin{equation}
\phi=-35^\circ, ~~~~~~~~\phi=+55^\circ.
\end{equation}
The mixing angle is unexpectedly large, and this is due to the
observed largeness of the leptonic decay width of $\psi(4160)$
(almost equal to that of the $\psi(4040)$). In fact, if  we neglect
the contribution from the 2D component of the $\psi(4160)$, we would
get an estimate for the mixing angle that is independent of
potential model parameters: $\phi\approx\pm 45^\circ$, which would
be the maximum mixing. The large 3S-2D mixing is a puzzling problem
in understanding the nature of $\psi(4160)$. Other studies like the
strong decays to $D^{*}\bar{D^{*}}$ may be useful to clarify the
3S-2D mixing problem for the $\psi(4160)$ (see, e.g. discussions in
\cite{barnes06}).

Despite of the above uncertainty concerning the 3S-2D mixing, we may
have a quite reasonable estimate of the production cross section of
$e^+e^-\to J/\psi+ \psi(4160)$, as compared with that of $e^+e^-\to
J/\psi+ J/\psi$. In the nonrelativistic limit, the charmonium masses
are all approximately set to be $M=2m_c$ (i.e. all binding energies
are neglected), and then we will have a simple relation
\begin{equation}
\frac{\sigma(e^+e^-\to J/\psi+ \psi(4160))}{\sigma(e^+e^-\to
J/\psi+J/\psi)}=\frac{\Gamma_{ee}(\psi(4160)}{\Gamma_{ee}(J/\psi)}\approx
0.15,
\end{equation}
where the observed values $\Gamma_{ee}(J/\psi)=5.55\pm 0.14\pm 0.02$
and $\Gamma_{ee}(\psi(4160))=0.83\pm $0.07~KeV \cite{pdg} are used.
This relation is obtained by the observation that in the double
vector-charmonium production via two virtual photons in $e^+e^-$
annihilation at $\sqrt{s}=10.6$~GeV  the photon fragmentation is
dominant (see e.g. \cite{cc,liuR}), in which the virtual photon
converts directly into the vector-charmonium, the same way as the
leptonic decay of the vector-charmonium.  As the most favorable
mechanism with the minimal photon-virtuality, all vector charmonium
states (e.g. $J/\psi, \psi(2S), \psi(4040), \psi(4160),...)$ are
expected to be produced from the two photon fragmentation in
$e^+e^-$ annihilation at $\sqrt{s}=10.6$~GeV or even higher
energies.

In ref.\cite{Pakhlov}, the following upper bound is given
\begin{equation}
\sigma(e^+e^-\to J/\psi+J/\psi)\times B(J/\psi\to >2~ charged)<
9.1~fb,
\end{equation}
which will imply
\begin{equation}
\sigma(e^+e^-\to J/\psi+\psi(4160)\times B(\psi(4160)\to > 2
~charged)< 1.4~fb,
\end{equation}
assuming $B(\psi(4160)\to > 2 ~charged)$ is comparable to
$B(J/\psi\to >2~ charged)$.

This predicted cross section is much smaller than the experimental
value given in eq.(2). Therefore,  the $\psi(4160)$ assignment for
the X(4160) should be ruled out.

\vspace{0.5cm}

III. The X(4160) could be an excited $0^{-+}$ charmonium state: the
$\eta_c(4S)$ (less likely to be the $\eta_c(3S)$).

As a possible candidate of the $0^{-+}$ state, the X(4160) can be
the $\eta_c(4S)$ charmonium, which is expected to decay into
$D^*\bar{D^*}$ and $D^*\bar{D}+c.c.$, but not $D\bar{D}$.

Note that Belle already found a new state, the X(3940), in the
process $e^++e^-\rightarrow J/\psi+X(3940)$\cite{belleX3940}, which
has a dominant decay mode into $D^*\bar D$ (with the fraction of
X(3940) decays with more than two charged tracks in the final state
into $D^*\bar D$ being $(96^{+45}_{-32}\pm 22)\%$), and a quite
narrow width $\Gamma=39\pm 26~MeV$. This result has been further
confirmed by Belle (see \cite{belle0707}. The X(3940) is considered
as a good candidate for the $\eta_c(3S)$ (for discussions see, e.g.
\cite{ELQ,swanson06,zhu07}). The problem is the low mass of X(3940)
as the $\eta_c(3S)$, compared with the $\psi(3S)$ candidate
$\psi(4040)$. But this could be explained by the coupled channel
effects that the coupling of $\eta_c(3S)$ to the $0^+$ and $0^-$
charmed meson pair (in S-wave) will lower the mass of
$\eta_c(3S)$\cite{ELQ}.

If we accept X(3940) as the $\eta_c(3S)$, then X(4160) should be the
$\eta_c(4S)$ if it is a $0^{-+}$ charmonium.  In this case, the mass
difference between $\eta_c(4S)$ and $\eta_c(3S)$ would be only
220~MeV . This mass difference is smaller than that predicted by the
potential models with linear plus Coulomb potentials (see, e.g.
\cite{Eichten:1979ms,Godfrey:1985xj,BGS0505002,ELQ}). Note that the
corresponding mass difference between the $\psi(4S)$ and $\psi(3S)$
is about 375~MeV if the $\psi(4S)$ is  identified with the
$\psi(4415)$ and the $\psi(3S)$ with the $\psi(4040)$ as
conventionally classified in the charmonium spectrum. An even more
puzzling problem is the mass splitting between the $\eta_c(4S)$ (if
identified with X(4160)) and the $\psi(4S)$ (if identified with
$\psi(4415)$), which is as large as 255~MeV, compared with the mass
differences between  $\eta_c(1S)$ and $J/\psi(1S)$,  $\eta_c(2S)$
and $\psi(2S)$, $\eta_c(3S)$ and $\psi(3S)$, which are only 117, 48,
and 100 Mev respectively (assuming the X(3940) is identified with
$\eta_c(3S)$). In simple potential models the mass splittings
between $0^{-+}(nS)$ and $1^{--}(nS)$ (n=1,2,3,4,...) are expected
to be decreased as $n$ increases. Although the mass spectrum can be
modified by the coupled channel effects and S-D mixing, such a big
mass difference, 255~MeV, between $\eta_c(4S)$ and $\psi(4S)$ is
still difficult to understand, unless the assignments for excited
$1^{--}$ states are changed in some way. For instance, if the
$\psi(4415)$ is not identified with the $\psi(4S)$ but with the
$\psi(5S)$, as discussed in the potential model with color screening
effects (see, e.g. \cite{screen93, screen95}), then the
corresponding $\eta_c(4S)$ mass could be lowered. In this case, all
higher excited states will be lowered in the mass spectra. But this
is only a plausible resolution for the problem in the $\eta_c(4S)$
assignment of X(4160), other approaches apparently need to be
studied.

Could the X(4160) be the $\eta_c(3S)$? If so, what assignment will
be for the X(3940). Moreover, if so, as the $\eta_c(3S)$ the mass of
X(4160) would be higher than that of $\psi(3S)$, which is identified
with $\psi(4040)$, by 120~MeV. The positive and large mass splitting
between $0^{-+}(3S)$ and $1^{--}(3S)$ seems not acceptable in
charmonium spectrum. So, X(4160) can not be the $\eta_c(3S)$.

The $\eta_c(4S)$ interpretation for the X(4160) is a likely one in
view of the large production rates of $\eta_c(1S)$, $\eta_c(2S)$,
and  $\eta_c(3S)$ (if identified with the X(3940)) associated with
$J/\psi$ in $e^+e^-$ annihilation. For the $\eta_c(4S)$ production,
the angular distribution and cross section is shown in Table I.
Note that the normalization can be substantially enhanced with the
NLO correction (see \cite{zhang}) but the angular distribution
remains unchanged in this case. The form of $(1+cos^2\theta)$ for
the angular distribution of this assignment differs markedly from
another interesting assignment, the $0^{++}$ charmonium (i.e. the
$\chi_{c0}(3P)$ state), which will be discussed below.

IV. The X(4160) might be an excited $0^{++}$ charmonium state: the
$\chi_{c0}(3P)$  (unlikely to be the $\chi_{c0}(2P)$).

As a possible candidate of the $0^{++}$ state, the X(4160) might be
the $\chi_{c0}(3P)$ charmonium, which is expected to decay into
$D^*\bar{D^*}$ and $D\bar{D}$, but not $D^*\bar{D}+c.c.$. The
$D\bar{D}$ decay mode of the X(4160) has not yet been seen so far.
The mass of X(4160) immediately indicates that it is unlikely to be
the $\chi_{c0}(2P)$ state, which is predicted to lie around
3.9-4.0~GeV. The fact that the observed Z(3930) can be identified
with the $\chi_{c2}(2P)$ state\cite{pdg}) also implies that the
$\chi_{c0}(2P)$ should lie well below 4160~MeV. So, X(4160) can only
be the $\chi_{c0}(3P)$ if it is a $0^{++}$ charmonium state.
However, if in $e^++e^-$ annihilation both $J/\psi+\chi_{c0}(1P)$
and $J/\psi+\chi_{c0}(3P)$ are observed, why $ J/\psi+\chi_{c0}(2P)$
is in the absence? In fact, according to the NRQCD calculation, the
production cross section of $ J/\psi+\chi_{c0}(2P)$ should be
comparable to or even larger than that of $ J/\psi+\chi_{c0}(1P)$
(see \cite{liu04}), because the first derivative of the wavefunction
at the origin for the 2P-state is usually larger than that for the
1P-state: $|R_{2P}^{'}(0)|^2> |R_{1P}^{'}(0)|^2$ (see,
e.g.\cite{wf}). To LO in NRQCD the cross sections for
$e^++e^-\rightarrow J/\psi+\chi_{c0}(1P)$ and $e^++e^-\rightarrow
J/\psi+\chi_{c0}(2P)$ are predicted to be 6.9~fb and 9.4~fb
respectively (QED contributions are included)~\cite{liu04}. So, the
experimental absence of $e^++e^-\rightarrow J/\psi+\chi_{c0}(2P)$
would be hard to understand.

However, at this point, it is interesting to notice that Belle has
observed a broad peak (but with only 3.8 $\sigma$) around
3.8-3.9~GeV in the recoil mass of $D\bar D$ against $J/\psi$ in the
process $e^++e^-\rightarrow J/\psi+D+\bar D$\cite{belle0707} (it may
also be seen in the $\gamma\gamma \to D\bar D$ process). Is this the
missing $\chi_{c0}(2P)$ state? If the bump in the 3.8-3.9~GeV region
is really due to the $\chi_{c0}(2P)\to D\bar D$ decay , the
$\chi_{c0}(3P)$ assignment for X(4160) would be favored (but the
$\chi_{c0}(2P)$ state should be further examined experimentally).

As discussed so far, two likely assignments for the X(4160) are the
$\eta_c(4S)$ and $\chi_{c0}(3P)$ charmonia. How to distinguish
between them? One effectible way is to measure the angular
distribution of the cross sections. The differential cross section
in the case of $\chi_{c0}(3P)$ is shown in Table I (see also
\cite{liu04}).
Compared with $(1+cos^2\theta)$ in the case of $\eta_c(4S)$, the
form of $(1+0.252cos^2\theta)$ for the  $\chi_{c0}(3P)$ has a much
weaker $\theta$ dependence,  and therefore the measurements on
angular distributions can be used to test the two possible
assignments. \vspace{0.5cm}

V. The X(4160) is unlikely to be the excited  $2^{++}$ or $1^{++}$
charmonium state, $\chi_{c2}(2P,3P)$ or $\chi_{c1}(2P,3P)$.

Experimentally, both $e^++e^-\rightarrow J/\psi+\chi_{c2}(1P)$ and
$e^++e^-\rightarrow J/\psi+\chi_{c1}(1P)$ have not been seen. This
is consistent with the smallness of calculated cross sections for
them. In fact to LO in NRQCD the cross sections for
$J/\psi+\chi_{c2}(1P)$ and $J/\psi+\chi_{c1}(1P)$ are predicted to
be 1.8~fb and 1.0~fb respectively (QED contributions are
included)~\cite{liu04}, which are much smaller than that for
$J/\psi+\chi_{c0}(1P)$ and $J/\psi+\eta_c(1S)$. In contrast, the
observed cross section for $J/\psi+X(4160)$ is comparable to that of
$J/\psi+\chi_{c0}(1P)$ and $J/\psi+\eta_c(1S)$. In view of both the
experimental non-observation and the calculated smallness of the
cross sections for $e^++e^-\rightarrow J/\psi+\chi_{c2}(1P)$ and
$e^++e^-\rightarrow J/\psi+\chi_{c1}(1P)$ we conclude that the
X(4160) is unlikely to be the excited $2^{++}$ or $1^{++}$
charmonium state, $\chi_{c2}(2P,3P)$ or $\chi_{c1}(2P,3P)$.

Note that the $\chi_{c2}(2P,3P)$ and $\chi_{c1}(2P,3P)$
interpretations for X(4160) are disfavored by the experimental
absence of $\chi_{c1,2}(1P)$ not only in the exclusive double
charmonium production of $\chi_{c1,2}(1P)$ associated with $J/\psi$,
but also in the inclusive prompt production of $\chi_{c1,2}(1P)$ in
$e^+e^-$ annihilation\cite{babar07}. In fact, recently Babar finds
no evidence for prompt $\chi_{c1,2}(1P)$ production after
subtracting the contributions from prompt $\psi(2S)$ production
feed-down to $\chi_{c1,2}(1P)$\cite{babar07}. Therefore, the
assignments of X(4160) as  $\chi_{c2}(2P,3P)$ or $\chi_{c1}(2P,3P)$
states are very unlikely.

\vspace{0.5cm}

VI. Non-charmonium interpretations for the X(4160): glueballs,
hybrids, and charmonium-molecules.

As suggested in \cite{bro}, a $0^{++}$ glueball associated with the
$J/\psi$ may be produced with a sizable rate in $e^+e^-$
annihilation. However, as a $0^{++}$ glueball, the X(4160) would
have a too large mass. Moreover, the glueball should mainly decay to
light hadrons, but not $D^*\bar{D^*}$. Nevertheless, to measure the
production angular distribution parameter $\alpha$, where the
differential cross section is proportional to $(1+\alpha~
cos^2\theta)$, will be useful to clarify the glueball interpretation
(with a negative value of $\alpha$ for the $0^{++}$ glueball).

Could the X(4160) be an exotic charmonium-hybrid? say, the $1^{-+}$
hybrid, a possible partner of the $1^{--}$ charmonium-hybrid $c\bar
cg$ candidate, the Y(4260)\cite{zhu,close}. However, the problem is,
a hybrid does not seem to have a favorable production mechanism
associated with the $J/\psi$ in $e^+e^-$ annihilation. Compared with
the production of double charmonium states such as
$e^++e^-\rightarrow J/\psi+\eta_c(1S)$, the production of a $c\bar
cg$ hybrid associated with $J/\psi$ requires an additional gluon
production, and could therefore be relatively suppressed. But
experimentally, the production rate of $X(4160)J/\psi$ is comparable
to that of $\eta_c(1S)J/\psi$.

Whether the X(4160) can be a charmonium molecule? The well know
charmonium-like state X(3872) has been suggested being a $D^0\bar
D^{*0}+c.c.$ molecule either as a real bound state (see, e.g.
\cite{braaten07} and references therein) or as a virtual state (see,
e.g.\cite{hanhart07}). The most significant motivation for the
molecule assignment is that the mass of X(3872) is very close to the
$D^0\bar D^{*0}$ threshold. However, in the case of X(4160), its
mass is above the $D^*\bar D^*$ threshold by about 140~MeV. This
makes the X(4160) unlikely to be a $D^*\bar D^*+c.c.$ molecule,
since for molecules the binding energies due to meson exchanges are
much less than 100~MeV.

\vspace{0.5cm}

In summary, we have discussed various interpretations of the
X(4160), observed by Belle in the process of double charm production
$e^+e^-\to J/\psi+X(4160)$ followed by $X(4160)\to D^*\bar{D^*}$.
The available information for this state from the data is its mass,
width, a major decay mode, and its large production rate (comparable
to $\eta_c(1S)$) associated with $J/\psi$ in $e^+e^-$ annihilation.

Using the leading order NRQCD calculation of the relative cross
sections of double charmonium production in $e^+e^-$ annihilation as
a guide, we  find that though the D-wave spin-singlet charmonium
state $^1D_2(2D)$ (with $J^{PC}=2^{-+}$) could be a candidate of
X(4160), the calculated production rate is too small, only about 5\%
of that for $e^+e^-\to J/\psi+\eta_c(1S)$, and therefore this
assignment is unlikely.

The possibility of X(4160) being the known $\psi(4160)$ produced via
two-photon fragmentation in $e^+e^-$ annihilation is also discussed,
but the calculated rate is much smaller (even with the 3S-2D mixing
effect included) than that for $e^+e^-\to J/\psi+J/\psi$ which,
however, only has a small experimental upper limit. So the
$\psi(4160)$ interpretation for X(4160) should be completely ruled
out.

The X(4160) is unlikely to be the excited  $2^{++}$ or $1^{++}$
charmonium state, e.g., $\chi_{c2}(2P,3P)$ or $\chi_{c1}(2P,3P)$,
because of the experimental non-observation and the calculated
smallness of the cross sections for the $J/\psi+\chi_{c2}(1P)$ and
$J/\psi+\chi_{c1}(1P)$ production.

The candidates of glueballs, $c\bar cg$ hybrids, and
charmonium-molecules for the X(4160) might also be considered, but
these interpretations are not very likely.

In contrast to above interpretations, the $\eta_c(4S)$ assignment
for X(4160) is an interesting possibility. The production rate of
$e^+e^-\to J/\psi+\eta_c(4S)$ relative to $e^+e^-\to
J/\psi+\eta_c(1S)$ in NRQCD is not very small, and could be
compatible with the Belle data (note that for the observed
$e^+e^-\to J/\psi+\eta_c(1S)$ cross section only a lower bound of
$25.6\pm 2.8\pm 3.4$~fb is given by Belle). And the non-observation
of the $D\bar D$ mode of X(4160) can also be understood for this
assignment. But one has to understand why $\eta_c(4S)$ has such a
low mass. And if one accepts X(4160) being the $\eta_c(4S)$, then
the $\psi(4415)$ can hardly be the $\psi(4S)$ as conventionally
classified in charmonium spectrum.

The $\chi_{c0}(3P)$  is an even more interesting candidate for the
X(4160). In particular, if the observed broad peak around
3.8-3.9~GeV in the recoil mass of $D\bar D$ against $J/\psi$ in
$e^++e^-\rightarrow J/\psi+D+\bar D$ \cite{belle0707} is due to the
$\chi_{c0}(2P)$ state, then the $\chi_{c0}(3P)$ assignment for
X(4160) will be favored.  However, as in the case of $\eta_c(4S)$
discussed above, one has to understand the problem of low mass
values of the 3P states in the $\chi_{c0}(3P)$ assignment, compared
with conventional potential model calculations. We also emphasize
that measurements on the angular distributions of cross sections are
useful to distinguish between the $\chi_{c0}(3P)$ and $\eta_c(4S)$
assignments for the X(4160).

In order to clarify the nature of  X(4160), it will be helpful
experimentally to measure the differential cross sections (the
production angular distributions), which can be different in
different assignments, and to measure the strong decay branching
ratios into various charmed meson pairs, and to measure the quantum
numbers of X(4160). Theoretically, it is certainly important to have
a reliable calculation for the strong decay rates, which is not very
easy considering the complexity due to the coupled channel effects,
and to understand why X(4160) has a dominant decay mode into
$D^*\bar{D^*}$. As for the production, as far as NRQCD is concerned,
the NLO radiative corrections are only available for the $e^+e^-\to
J/\psi+\eta_c$ process\cite{zhang}, which is confirmed by a recent
independent calculation\cite{wang}. It will certainly be very useful
if the NLO calculation for the P-wave and D-wave charmonium states
involved in double charmonium production can be performed.

At present, we only have a limited understanding for the puzzling
state X(4160). Since its finding was reported more than five months
ago, there have been no theoretical papers on its interpretations.
So, it is our hope that the discussion presented in this paper will
stimulate more interesting discussions on this new hadronic state.


\vspace{0.5cm}

$Acknowledgments.$~~The author would like to thank C.F. Qiao, C.Z
Yuan and S.L. Zhu for helpful discussions and comments. He also
thanks T. Barnes and Q. Zhao for a general discussion on charmonium
spectroscopy. This work was supported in part by the National
Natural Science Foundation of China (No. 10675003, No 10721063), and
the Research Found for Doctorial Program of Higher Education of
China.

\newpage
\begin{center}
\begin{table*}
\caption{Angular distributions and cross sections for double
charmonium production in $e^+ e^-$ annihilation at
$\sqrt{s}=10.6$~GeV with both QCD and QED contributions in the
leading order NRQCD calculations (numbers without QED contribution
are given with square brackets, see also text for the input
parameters).}

\begin{tabular}{c|cc}
\hline
  Final state & Differential cross section (fb)& Cross section  (fb) \\\hline
  $J/\psi + \eta_{c}(1S)$ & $2.47~ [2.06](1+cos^2\theta)$ & 6.6 ~[5.5] \\\hline
  $J/\psi + ^1D_2(1D)$ & $0.077 ~[0.069](1+cos^2\theta)$ & 0.21 ~[0.19] \\\hline
  $J/\psi + ^1D_2(2D)$ & $0.123 ~[0.111](1+cos^2\theta)$ & 0.34 ~[0.31] \\\hline
  $J/\psi + \eta_{c}(4S)$ & $1.14~ [0.95](1+cos^2\theta)$ & 3.0~ [2.5] \\\hline
  $J/\psi + \chi_{c0}(3P)$ & $4.7 ~[4.6](1+0.252cos^2\theta)$ & 10.2 ~[9.9] \\\hline

\end{tabular}
\end{table*}
\end{center}

\end{document}